\documentclass[apjpt4]{aastex}

\newcommand \msun{$\rm M_\odot$}

\newcommand\nuk[2]{$\rm ^{\rm #2} #1$}

\shortauthors{Chieffi,Limongi}
\shorttitle{The synthesis of \nuk{Ti}{44} and \nuk{Ni}{56}}
\received{October 28, 2016}
\accepted{January 09, 2017}
\slugcomment{Submitted to The Astrophysical Journal - accepted January 9 2017}

\begin{document}

\title{The synthesis of \nuk{Ti}{44} and \nuk{Ni}{56} in massive stars}
\author{Alessandro Chieffi\altaffilmark{1,3} and Marco Limongi\altaffilmark{2,3,4}}

\altaffiltext{1}{Istituto Nazionale di Astrofisica - Istituto di Astrofisica e Planetologia Spaziali, Via Fosso del Cavaliere 100, I-00133, Roma, Italy;
alessandro.chieffi@inaf.it}

\altaffiltext{2}{Istituto Nazionale di Astrofisica - Osservatorio Astronomico di Roma, Via Frascati 33, I-00040, Monteporzio Catone, Italy;
marco.limongi@oa-roma.inaf.it}

\altaffiltext{3}{Centre for Stellar \& Planetary Astrophysics, 
School of Mathematical Sciences, P.O. Box, 28M, Monash University, Victoria 3800, Australia}

\altaffiltext{4}{Kavli Institute for the Physics and Mathematics of the Universe, Todai Institutes for Advanced Study, the University of Tokyo, Kashiwa, Japan 277-8583 (Kavli IPMU, WPI)}              	

\begin{abstract}
We discuss the influence of rotation on the combined synthesis of \nuk{Ti}{44} and \nuk{Ni}{56} in massive stars. While \nuk{Ni}{56} is significantly produced by both the complete and incomplete explosive Si burning, \nuk{Ti}{44} is mainly produced  by the complete explosive Si burning, with a minor contribution (in standard non rotating models) from the incomplete explosive Si burning and the O burning (both explosive and hydrostatic). We find that, in most cases, the thickness of the region exposed to the incomplete explosive Si burning increases in rotating models ($\rm v_{\rm ini}$=300 km/s) and since \nuk{Ni}{56} is significantly produced in this zone, the fraction of mass coming from the complete explosive Si burning zone, necessary to get the required amount of \nuk{Ni}{56}, reduces. Therefore the amount of \nuk{Ti}{44} ejected for a given fixed amount of \nuk{Ni}{56}  decreases in rotating models. However, some rotating models at [Fe/H]=-1 develop a very extended O convective shell in which a consistent amount of \nuk{Ti}{44} is formed, preserved and ejected in the interstellar medium. Hence a better modeling of the thermal instabilities (convection) in the advanced burning phases together with a critical analysis of the cross sections of the nuclear reactions operating in O burning are relevant for the understanding of the synthesis \nuk{Ti}{44}.
\end{abstract}
\keywords{stars: abundances, stars: evolution, stars: interiors, stars: massive, stars: rotation, supernovae: general}

\section{Introduction} \label{sec:intro}
Neutral \nuk{Ti}{44} is an isotope unstable to $\rm e^-$ capture with a half life of 58.9 $\pm$ 0.3 yr \citep{ah06}. It decays to \nuk{Sc}{44} first by emitting a $\gamma$ of 1157 keV and to \nuk{Ca}{44} later by emitting two additional $\gamma$ rays of 67.9 and 78.4 keV. Since the 60's \citep{bcf68,wac73} it has been recognized that it may be produced in the very deep regions of a massive star during the explosion, if they are shocked to very high temperatures (greater than 5 GK or so) to reach Nuclear Statistical Equilibrium but then cooled (expanded) rapidly enough that a large amount of  free $\alpha$ particles is left ($\alpha$ rich freeze-out).  In this case, in fact,  it exists a temporal "window" of the order of 200/300 ms in which the local temperature T and density $\rho$  of these expanding layers are sufficiently high for the nuclear reactions to activate in presence of fuel ($\alpha$ particles in this case). Within this scenario the synthesis of \nuk{Ti}{44} has been always explored by parametric studies of the properties of the $\alpha$ rich freeze-out as a function of various parameters, mainly T, $\rho$, electron mole number Ye and the relevant nuclear reaction rates \citep{the98,the06,mth10}. An important constraint a model must satisfy to provide a meaningful prediction of \nuk{Ti}{44}, is to  avoid overproduction of \nuk{Ni}{56}, another unstable nucleus synthesized in complete and incomplete explosive Si burning.

From an observational point of view the quest for a signal from the decay of  \nuk{Ti}{44} started as soon as the first X- and $\gamma-$ ray detectors where launched in the 80's \citep{mwh92,ls94,db97}. After more than 30 years of data taken by several satellites, at present we have only two clear evidences of  the presence of  live \nuk{Ti}{44}: a first one from the supernova remnant Cas A and a second one from the SN1987A in the Large Magellanic Cloud. 

The signal from Cas A is well secured and the latest available data (the 78.36 and 1157 keV lines detected by INTEGRAL give $1.37 \pm 0.19 \times10^{-4}$ \msun ~- \cite{sdk15}; the 67.86 and 78.36  keV lines detected by NuSTAR give $1.25 \pm 0.3 \times10^{-4}$ \msun ~- \cite{ghb14}) converge towards an amount of \nuk{Ti}{44} of the order of $1\div1.3\times10^{-4}$ \msun. Another recent finding  \citep{ghb14} concerns the strong asymmetries in the spatial distribution of \nuk{Ti}{44} around this supernova remnant together with the fact that it appears uncorrelated with the Fe X-ray emission (CHANDRA data). Though these data are fundamental and necessary to constrain the explosion properties of this star, unfortunately a reliable estimate of the amount of \nuk{Ni}{56} ejected is missing. According to the analysis of the proper motion of the ejecta, this supernova should have exploded in 1671 but none reported the appearance of a "new" star in those years (with the possible exception of  Flamsted in 1680). Since the luminosity of a supernova is directly connected to the amount of \nuk{Ni}{56} ejected during the explosion (because the light curve is powered by the decay of \nuk{Ni}{56} first and \nuk{Co}{56} later) and given its proximity (3.4 kpc), the lack of detection put strong limits on the maximum amount of \nuk{Ni}{56} ejected. The situation is unfortunately even more complex because the explosion could have been obscured by the presence of a large amount of circumstellar matter. A recent analysis of the reddening in the direction of this supernova remnant \citep{ea09} shows that the amount of \nuk{Ni}{56} could have been as large as 0.15 \msun ~and still be not visible from the Earth. At present we can only state that we do not know how much \nuk{Ni}{56} was ejected in this event.

As far as SN1987A is concerned, we have certainly more stringent data since we know both the amount of \nuk{Ni}{56} ejected during the explosion ($\simeq$0.075 \msun; \cite{cwf88,stm14}) and the amount of \nuk{Ti}{44} present in the ejecta. In fact, NuSTAR detected both the 67.86 and 78.36  keV lines and derived a \nuk{Ti}{44} abundance of $1.5 \pm 0.3 \times10^{-4}$ \msun ~- \cite{bhm15} -while INTEGRAL was able to measure only a combined flux of  the 67.86 and 78.36  keV lines and derived a \nuk{Ti}{44} abundance of $3.1 \pm 0.8 \times10^{-4}$ \msun ~- \cite{glt12}. A recent analysis of the UVOIR light curve of 1987A by \cite{stm14} reported an amount of \nuk{Ti}{44} of the order of $0.55 \pm 0.17 \times10^{-4}$ \msun~ necessary to power the light curve at late times. Given the uncertainties (that are larger than the formal error bars) we feel confident to say that very probably 1987A ejected something between 1 and 2 $\times10^{-4}$ \msun ~of \nuk{Ti}{44}. The somewhat poorer knowledge of the amount of \nuk{Ti}{44} ejected in this explosion is largely counterbalanced by a very good knowledge of the amount of \nuk{Ni}{56} ejected.

\begin{figure}[h]
\center{\includegraphics[scale=0.20]{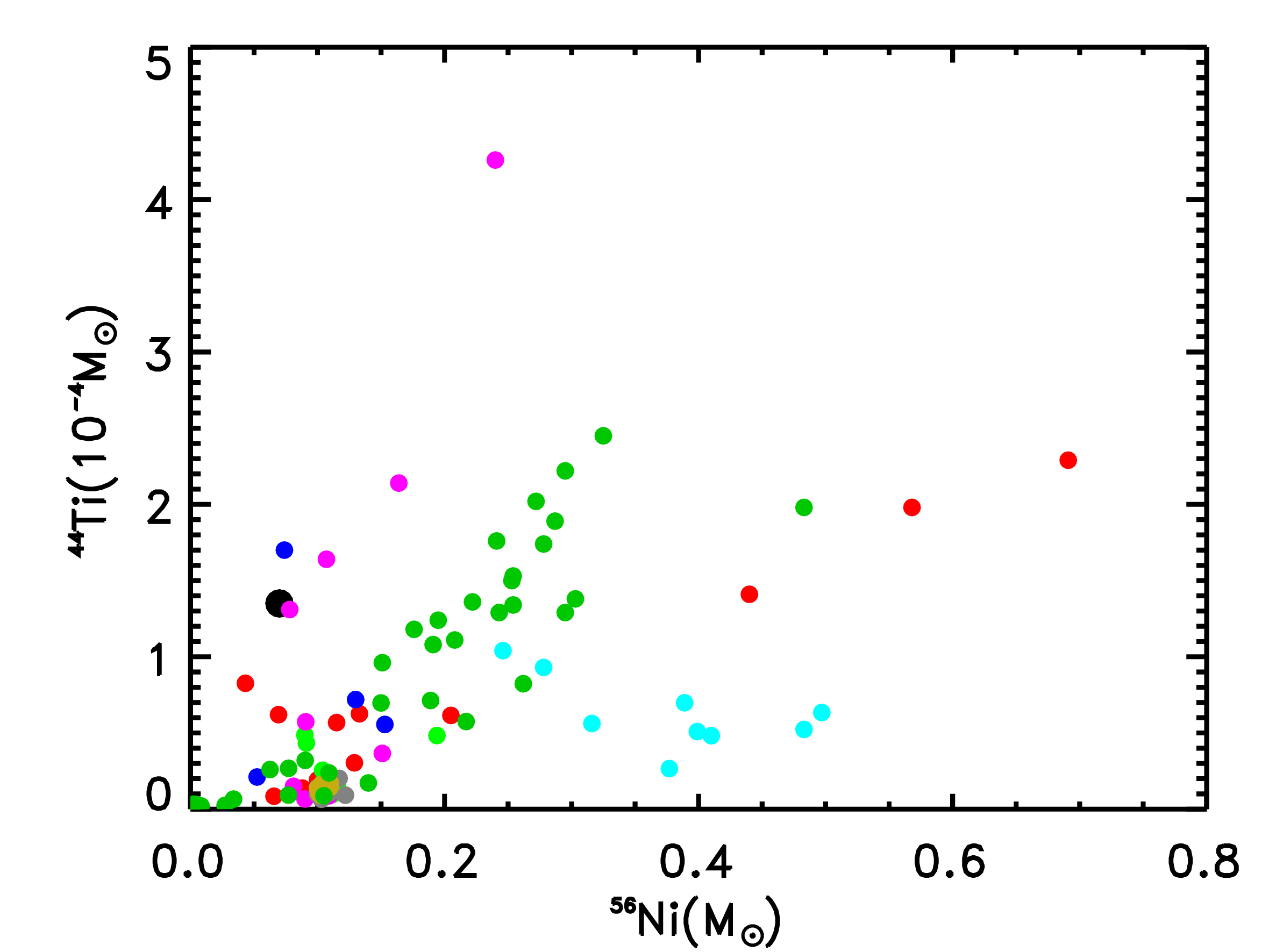}}
\caption{Ejected amount of \nuk{Ti}{44} versus \nuk{Ni}{56} for different sets of models;  red \citep{ww95}, blue \citep{tnh96}, magenta \citep{mn03}, cyan \citep{mth10}, light green \citep{rhf02}, dark green \citep{lc03}, brown and grey \citep{cl12} non rotating and rotating models, respectively. The black dot represents the position of the SN 1987A. \label{ti44vni56}}
\end{figure}

As far as we know, available evolutionary models fail to predict the right amount of \nuk{Ti}{44} (corresponding to an amount of \nuk{Ni}{56}$\simeq$0.07 \msun) by a factor of the order of 3 or more. Figure \ref{ti44vni56} shows the amount of \nuk{Ti}{44} and \nuk{Ni}{56} ejected by a number of models (see the figure caption). The big black dot marks the values corresponding to the supernova remnant 1987A.   It is quite evident that basically no models are compatible with the observed values. The only one that fits 1987A comes from an aspherical explosion of a pure He core of 8 \msun ~that should represent the He core of a 25 \msun~ \citep{mn03}. The interpretation  of Cas A is more difficult but it is clear that the observed amount of  \nuk{Ti}{44} could be reconciled with the existing models only if the exploding star would have ejected at least 0.15 \msun ~of \nuk{Ni}{56}.

After the publication of a first set of rotating solar metallicity models \citep{cl12}, we have now completed the computation of a much larger set of models that extends in mass between 13 and 120 \msun, in metallicity between [Fe/H]=0 and -3 and covers three initial rotational velocities (0, 150 and 300 km/s). All the details of these new models will be published in a companion paper \citep{lc16}. Here we have extracted from that large set the yields of \nuk{Ti}{44} and \nuk{Ni}{56} because we think they worth a separate discussion. This paper is organized as follows: the basic properties of the models are reported in the next section while an analysis of the results is presented in section \ref{disc}.

\section{The models}\label{models}
The results presented in this paper are based on a grid of models having initial masses 13, 15, 20, 25, 30, 40, 60, 80 and 120 \msun, initial metallicities [Fe/H]=0, -1, -2, -3 and initial equatorial velocities v=0, 150, 300 km/s. The adopted solar chemical composition is the one by \cite{asp09}. At lower metallicities, following  \cite{caetal04} and \cite{spiteetal05}, we assume that a few elemental species are enhanced with respect to the scaled solar composition. In particular we adopt
[C/Fe]=0.18, [O/Fe]=0.47, [Mg/Fe]=0.27, [Si/Fe]=0.37, [S/Fe]=0.35, [Ar/Fe]=0.35, [Ca/Fe]=0.33 and [Ti/Fe]=0.23 at all metallicities lower than solar. The initial He abundances are Y=0.265 ([Fe/H]=0), 0.25 ([Fe/H]=-1) and 0.24 ([Fe/H]$<$-1).

All models were followed from the pre Main Sequence phase up to the onset of the final collapse by means of the latest version of our code, the FRANEC. The main features of this code, as well as all the input physics and assumptions, have been already extensively discussed in \cite{cl12} and will be not repeated here. The only improvements with respect to the version described in \cite{cl12} are: (1) a better treatment of the angular momentum transport in the envelope of the star; (2) the inclusion of a proper mass loss that activates when the star approaches the Eddington limit; (3) a refined computation of the angular momentum loss due to the stellar wind and (4) a more extended nuclear network. Though the evolutionary properties of all these stars will be discussed in \cite{lc16}, it is worth mentioning here that one of the key (and direct) effects of rotation on the advanced burning phases is a systematic reduction of  the \nuk{C}{12} $/$ \nuk{O}{16} ratio as a consequence of the continuous ingestion of fresh He during the latest phases of the central He burning where most of the conversion of \nuk{C}{12} to \nuk{O}{16} occurs. This is relevant, in the present context, because the lower the concentration of \nuk{C}{12} the faster the C burning shell advances in mass leaving room for the possible formation of an extended O convective shell (see below).

The explosion of the mantle of each stellar model was followed by means of a hydrodynamic code developed by us that solves the fully compressible reactive hydrodynamic equations using the  Piecewise Parabolic Method (PPM) of \citet{cw84} in lagrangean form. Since the explosions cannot be computed yet on the basis of first principles, we still have to rely on a parametric approach in which some arbitrary amount of energy is injected in the deep interior of the models. More specifically each explosion is started by means of a kinetic bomb, i.e. by imparting instantaneously an initial velocity $v_{0}$ to a mass coordinate of $\sim 1~M_\odot$, i.e. well within the iron core \citep{lc06} and followed for $10^8$ s, well after the expanding envelope has become homologous. It goes without saying that each (arbitrary) initial velocity  $v_{0}$ will correspond to a specific mass cut and abundances of all the nuclear species synthesized in the deepest regions of the star. The explosions presented in Table 1 were computed by requiring that each model ejects of 0.07 \msun ~of \nuk{Ni}{56}.

\section{Discussion}\label{disc}
\begin{figure}[h]
\center{\includegraphics[scale=0.20]{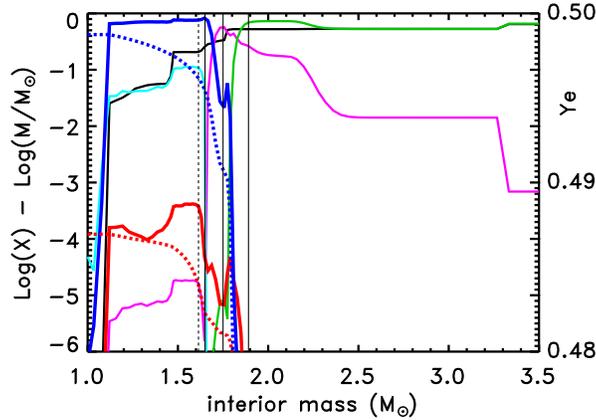}}
\caption{The distribution of \nuk{Ti}{44} and \nuk{Ni}{56} in a non rotating 20 \msun~ of solar metallicity after the passage of the shock wave.  The various lines refer to: O (green), Si (magenta), $\alpha$ (cyan), \nuk{Ti}{44} (red), \nuk{Ni}{56} (blue), Ye (black), integrated \nuk{Ti}{44} (dashed red), integrated \nuk{Ni}{56} (dashed blue). All abundances are in mass fractions except for the two integrated ones that are in solar masses. The 3 solid black vertical lines mark, left to right, the borders of the Six, Siix and Ox burning regions, while the dashed black vertical line marks the mass location corresponding to 0.07 \msun ~of \nuk{Ni}{56} ejected.\label{m20}}
\end{figure}

The profiles of the abundances (in mass fraction) of O, Si, $\alpha$, \nuk{Ti}{44} and \nuk{Ni}{56} after the passage of the shock wave, within the innermost layers of a non rotating 20 \msun~ star of solar metallicity are shown in Figure \ref{m20} together with the electron mole number Ye and the integrated (from the surface of the star) abundances (in solar masses) of the two unstable nuclei. The three black solid vertical lines mark, left to right, the mass coordinates of the outermost layers where complete explosive Si burning (Six), incomplete explosive Si burning (Siix) and explosive O burning (Ox) occur. The dashed black vertical line marks the mass location corresponding to 0.07 \msun~ of \nuk{Ni}{56} ejected. \nuk{Ti}{44} shows a major production in the Six region in presence of an $\alpha$ rich freeze-out (note the high final $\alpha$ abundance in this region) but it shows also the presence of two minor peaks, one in the region of the Siix and another one in the region of Ox. \nuk{Ni}{56} shows a composite production too, so that the relative abundance between these two nuclei depends on the region where they are synthesized. The cumulative abundance of both nuclei reflects such different behaviors. Moving towards the interior both show a first steep raise due to the production by the Ox, then a shallower raise that reflects the contribution of the Siix and eventually the final main raise due to the major contribution from the Six. 

\begin{figure}[h]
\center{\includegraphics[scale=0.20]{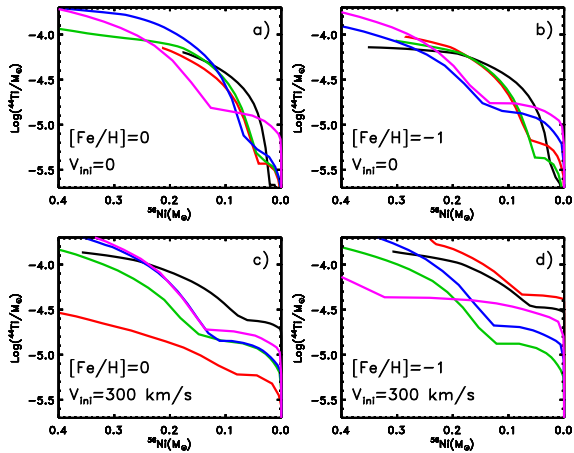}}
\caption{The trend of \nuk{Ti}{44} versus \nuk{Ni}{56} for a sample of stars. The various lines refer to: 13 \msun~ (black), 15 \msun~ (red), 20 \msun~ (green), 25 \msun~ (blue), 40 \msun~ (magenta).  The (left) end point of each line corresponds to the maximum amount of \nuk{Ni}{56} that may be ejected without ejecting simultaneously matter from the Fe core mass.} \label{runti44ni56}
\end{figure}

To visualize how the two cumulative abundances are connected one to the other, Figure \ref{runti44ni56} shows a plot of \nuk{Ti}{44} versus \nuk{Ni}{56} for a subset of solar metallicity non rotating models in panel a). Each line refers to a stellar model and each point along a given line represents the amount of \nuk{Ti}{44} that would be ejected together with the corresponding amount of \nuk{Ni}{56}. Though the general trend is that the \nuk{Ti}{44} ejected scales directly with \nuk{Ni}{56}, it is possible to recognize in Figure \ref{runti44ni56} the different production zones identified in Figure \ref{m20}. Each mass shows a first raise of \nuk{Ti}{44} when the amount of \nuk{Ni}{56} is still negligible (this component reflects the production by the Ox), then a shallow raise that corresponds to the contribution of the Siix and the final steep raise that marks the contribution of the Six.

Table 1 shows in the first four columns,  for each metallicity and initial rotational velocity (and an assumed yield of 0.07 \msun ~of \nuk{Ni}{56}, value close to the one determined for the SN1987A) the following quantities: the initial mass, the mass cut and the inner borders of the regions exposed, respectively, to the Siix and the Ox (all in solar masses). Column 5 to 10 show the integrated abundances of the two unstable nuclei (again in solar masses) at the three mass coordinates given in columns 2 to 4 , while the remaining columns show the respective percentage of production in the various zones. Note that when the mass cut falls in the region of the Siix, the quantities reported in the columns marked as Siix and Mcut obviously coincide. 

The first thing worth noting is that the amount of \nuk{Ti}{44} ejected by non rotating models of solar metallicity ranges between $\rm 0.1~and~ 0.3 \times10^{-4}$ \msun~ and falls short at least by a factor of four of the observed value of $\sim 1.3\times10^{-4}$ \msun~ (column 5 in the Table). Note that the mass cut falls within the Siix region in stars more massive than 25 \msun ~(see columns 2 and 3): this means that these stars do not eject any matter exposed to the Six (columns 11 and 14) region where the maximum production of \nuk{Ti}{44} occurs. Even in the range 15-25 \msun ~a significant fraction of the 0.07 \msun ~of \nuk{Ni}{56} does {\it not} come from the Six region. In other words, since a large fraction of the required amount of \nuk{Ni}{56} comes from layers more external to the region exposed to the Six, it is not possible to extract much matter from the Six zone where most of \nuk{Ti}{44} is made:  this explains why the yields of \nuk{Ti}{44} are very low. However, even if it were possible to extract only matter exposed to Six (cancelling the contributions of the more external regions, i.e. the Six and the Ox), the amount of \nuk{Ti}{44} that would correspond to 0.07 \msun ~of \nuk{Ni}{56} fully produced in the Six, would not exceed $\sim 4\times10^{-5}$ \msun. Panel a) in Figure \ref{runti44ni56} clearly shows that an amount of \nuk{Ti}{44} of the order of $10^{-4}$ \msun ~would require the ejection of more than 0.2 \msun ~of \nuk{Ni}{56}, a value too large with respect to the one observed in the SN1987A.

\begin{figure}[h]
\center{\includegraphics[scale=0.15]{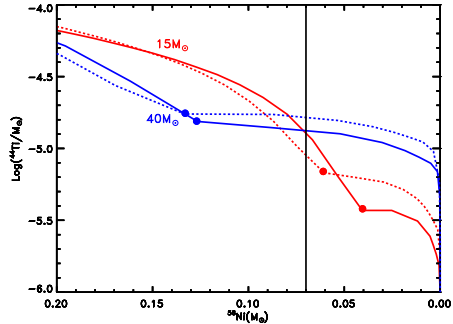}}
\caption{Comparison between the distribution of \nuk{Ti}{44} versus \nuk{Ni}{56} in two stars of [Fe/H]=0 (solid lines) and [Fe/H]=-1 (dashed lines). The red lines refer to a 15 \msun~ while the blue lines to a 40 \msun. The filled dots mark the passage from the Six to the Siix. The solid black vertical line marks the mass location corresponding to 0.07 \msun~ of \nuk{Ni}{56} ejected.   \label{confFe}}
\end{figure}

Lowering the metallicity does not help. Models computed for [Fe/H]=-1 do not vary significantly from the solar ones - third panel in Table 1 and panel b) in Figure \ref{runti44ni56} - because both  \nuk{Ti}{44} and  \nuk{Ni}{56} are primary elements and therefore depend on the metallicity only indirectly through its influence on the evolutionary properties of a star (e.g. size of convective core, mass loss) but do not have a direct dependence on the initial metallicity (like secondary elements: e.g. N and the s-processes). At this metallicity all stars more massive than 20 \msun ~produce more than 0.07 \msun ~of \nuk{Ni}{56} outside the region of the Six and hence also in this case it is difficult to extract material from the Six zone (even an amount of the order of 0.10 \msun ~of \nuk{Ni}{56} would leave this discussion unaltered). Moving from [Fe/H]=0 to [Fe/H]=-1,  the yield of \nuk{Ti}{44} reduces somewhat in the range 13 to 20 \msun, while it mildly increases in the more massive stars. Such a dependence must be considered with care because it largely depends on the adopted mass cut. Figure \ref{confFe} shows (for both a 15 and a 40 \msun) a comparison of the \nuk{Ti}{44} versus \nuk{Ni}{56} relation between the two metallicities. The red lines refer to the 15 \msun~ while the blue ones to the 40 \msun. The solid and dashed lines refer to [Fe/H]=0 and [Fe/H]=-1 respectively. The Figure clearly shows that the amount of \nuk{Ti}{44} (as a function of the \nuk{Ni}{56} ejected!) produced by the Ox and the Siix (right side of the filled dots) increases as the metallicity decreases, while the opposite occurs within the region of Six (left side of the filled dots). Actually there may be more than one intersection, due to the complex and non monotonic interplay among the Ye profile, the mass-radius relation and the passage of the shock wave. For the specific choice of 0.07 \msun~ of \nuk{Ni}{56} (solid black vertical line in Figure \ref{confFe}), the mass cut falls in the region of the Six for the mass range range 13 to 20 \msun~ and hence the \nuk{Ti}{44} scales inversely with initial metallicity. The more massive stars, vice versa, reach the chosen amount of \nuk{Ni}{56} in the Siix and hence they show a direct scaling with the initial metallicity. However what really matters is that the dependence on the metallicity is in any case quite modest, remaining within a factor of two or so for the range of \nuk{Ni}{56} of interest.

Summarizing the result obtained so far, analogously to what has been found for more than two decades by most authors working with 1D non rotating models and spherically symmetric explosions, we cannot explain the \nuk{Ti}{44} synthesized by the SN1987A. The analysis of Cas A is much less stringent due to the lack of a good determination of the \nuk{Ni}{56} ejected in that event (see the Introduction).

\begin{figure}[h]
\center{\includegraphics[scale=0.20]{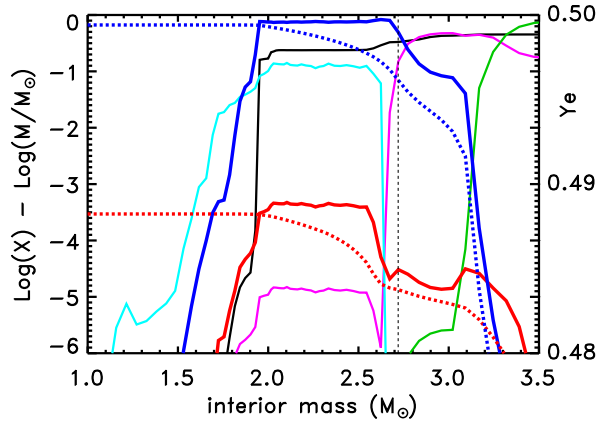}}
\caption{The distribution of \nuk{Ti}{44} and \nuk{Ni}{56} in a rotating 20 \msun~ of solar metallicity after the passage of the shock wave. The various lines refer to: O (green), Si (magenta), $\alpha$ (cyan), \nuk{Ti}{44} (red), \nuk{Ni}{56} (blue), Ye (black), integrated \nuk{Ti}{44} (dashed red), integrated \nuk{Ni}{56} (dashed blue). The dashed black vertical line marks the mass location corresponding to 0.07 \msun~ of \nuk{Ni}{56} ejected.   \label{m20r}}
\end{figure}

Rotation basically leads to more massive He cores and to a lower amount of C at central He exhaustion (see Section \ref{models}). Hence it primarily affects the Mass-Radius relation at the onset of the collapse and the amount of mass that will be exposed to the various explosive burning. As an example, Figure \ref{m20r} shows the structure of the rotating 20 \msun~ of solar metallicity after the passage of the shock wave. The regions exposed to the various explosive burning are clearly much more extended in mass with respect to the non rotating case (Figure \ref{m20}), the Siix region, for example, extending over roughly half a solar mass, a factor of four or so bigger than in the non rotating case. The second panel in Table 1 shows the data for the rotating solar metallicity  models while panel c) in Figure \ref{runti44ni56} shows the corresponding trend of \nuk{Ti}{44} versus \nuk{Ni}{56}. Rotating models at solar metallicity reach 0.07 \msun~ of \nuk{Ni}{56} in the Siix region in all models more massive than 15 \msun. The consequence on the yields of \nuk{Ti}{44} is that in most cases they are {\it even lower} than in the non rotating case. Note, however, that the amount of \nuk{Ti}{44} produced by the Ox increases because the thickness of the region exposed to this burning increases.

\begin{figure}[h]
\center{\includegraphics[scale=0.20]{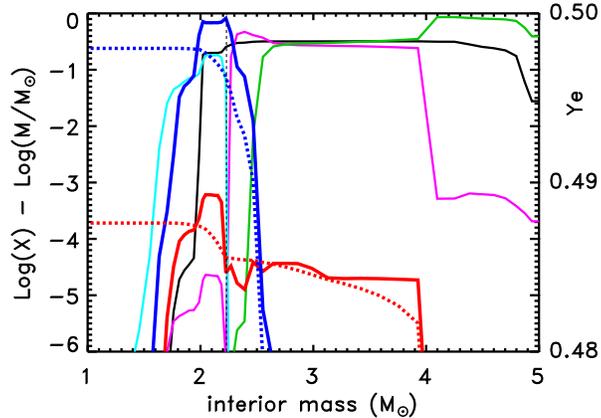}}
\caption{The distribution of \nuk{Ti}{44} and \nuk{Ni}{56} in a rotating 15 \msun~ of [Fe/H]=-1 after the passage of the shock wave. The various lines refer to: O (green), Si (magenta), $\alpha$ (cyan), \nuk{Ti}{44} (red), \nuk{Ni}{56} (blue), Ye (black), integrated \nuk{Ti}{44} (dashed red), integrated \nuk{Ni}{56} (dashed blue).  The dashed black vertical line marks the mass location corresponding to 0.07 \msun~ of \nuk{Ni}{56} ejected. \label{m15br}}
\end{figure}

Rotating models at [Fe/H]=-1 (summarized in the forth panel in Table 1 and panel d) in Figure \ref{runti44ni56}) show a qualitatively similar behavior, the rotating stars reaching again 0.07 \msun~ of \nuk{Ni}{56} well within the Siix region in all models more massive than 15 \msun. This time however, the amount of \nuk{Ti}{44} produced is larger than in the non rotating case because of a more consistent contribution of the Siix and Ox to its synthesis. But low metallicity rotating models show a very interesting feature in the two lowest masses, 13 and 15 \msun, i.e. the formation of a wide O convective shell that extends over more than 1.5 \msun. The O burning shell is always a nursery of \nuk{Ti}{44} - such an occurrence was already noted by \cite{tha10} - but this layer is always so close to the mass cut that it is completely swept out by the shock wave. The formation of a very extended O convective shell, on the contrary, preserves most of this \nuk{Ti}{44} because convection pushes the freshly made \nuk{Ti}{44} at a large distance where it is not affected by the passage of the blast wave. Figure \ref{m15br} shows the distribution of the nuclei relevant for the present discussion in a rotating 15 \msun~ having [Fe/H]=-1, and in particular the large amount of \nuk{Ti}{44} produced by the O burning shell spread over the wide convective region and left almost untouched by the shock wave. The formation of unusually extended convective shells is not so rare. Zero metallicity stars, for example, often experience such an occurrence due to the low entropy barrier between the He and H rich zones. We have already shown in \cite{lc06} that the extension of convective zones affects significantly the yields of other nuclear species, as \nuk{Al}{26} and \nuk{Fe}{60}. Let us also remark that the current description of convection in general, but in particular in the more advanced phases, may be very different from what we model in 1D: according to the studies of, e.g., \cite{ma07} and \cite{am09}, the O convective shell could develop through flames that could extend even very far from the burning location. The interest of this result is also due to the fact that it could account for a different behavior between the average stars, that do not produce a large amount of \nuk{Ti}{44}, and some specific cases in which rotation may lead to the formation of an extended convective shell where a much larger amount of \nuk{Ti}{44} may be synthesized and preserved. 

Another important constraint that cannot be neglected in the analysis of the combined synthesis of \nuk{Ti}{44} and \nuk{Ni}{56} is the ratio \nuk{Ca}{44} / \nuk{Fe}{56} in the solar chemical composition. According to \cite{asp09} this ratio (by number) in the Sun is $1.57\times10^{-3}$. By assuming that all solar metallicity stars between 13 and 120 \msun~ eject 0.07 \msun~ of \nuk{Ni}{56} and integrating over a Salpeter IMF (x=1.35) we get a ratio \nuk{Ca}{44} / \nuk{Fe}{56}$=7.1\times10^{-4}$ in the non rotating case and equal to $1.1\times10^{-3}$ for rotating models. In such a scenario the ejecta of stars more massive that 25 \msun~ must have a final kinetic energy in excess of $3\times10^{51}$ erg in order to eject even a minor fraction of \nuk{Ni}{56}, and this amount is incompatible the average kinetic energy of a sample of core collapse supernovae \citep{pp15,lyman16}; if one would fix the maximum kinetic energy of the ejecta to $3\times10^{51}$ erg, all stars more massive that 25 \msun~ fail to explode and collapse completely, contributing to the chemical enrichment  only through the wind. Note that also the works of \cite{oo11}  and \cite{suk16} support this idea: \cite{oo11} define, on the basis of a large set of hydro simulation, a compactness parameter $\xi$  that allows to "predict" the final fate of a collapsing star: they find that a value of $\xi$ of the order of 0.45 marks the passage from structure that collapse to a black hole from those who don't. All our models of mass larger than 25 \msun~ have a compactness parameter $\xi$ well above 0.45. \cite{suk16} followed the explosions of a  very fine grid of models in the range 9 to 120 \msun~ and found a complex, non continuous distribution of models that explode and don't explode: however most of their models having initial mass larger than 28 \msun~ or so fully collapse without exploding. In a scenario in which stars more massive than 25 \msun~ fail to explode, our models would predict these ratios (by number) for  the \nuk{Ca}{44}/ \nuk{Fe}{56}: {$1\times10^{-3}$([Fe/H]=0, v=0) and {$1.4\times10^{-3}$([Fe/H]=0, v=300 km/s). Given the many uncertainties in the modeling of both the hydrostatic evolution as well as the explosion of a stellar model, we think that the present set of models predicts a ratio compatible with the observed value, in particular for the second scenario in which all stars more massive than 25 \msun~ are assumed to completely collapse without any ejecta apart from the mass lost through the wind. 

All these results imply that the amount of \nuk{Ti}{44} and \nuk{Ni}{56} predicted by our models can account for the \nuk{Ca}{44}/\nuk{Fe}{56} ratio in the solar chemical composition without the necessity of additional \nuk{Ti}{44} from the majority of the stars. This results also agrees with the fact that, if any massive star would eject an amount of \nuk{Ti}{44} of the order of $1.3\div1.5\times10^{-4}$ \msun, the current all sky maps should have detected a clear signal from,  e.g., the galactic centre, that is not observed \citep{tk16}.

SN 1987A and possibly Cas A could be rarer events in which the formation of an extended O convective shell (due to rotation and/or a more reliable description of the thermal instabilities) could contribute to the synthesis of \nuk{Ti}{44}. A (partial) decoupling of the region where \nuk{Ti}{44} and \nuk{Ni}{56} are synthesized could also help in the understanding of the lack of correlation between the \nuk{Ti}{44} and the Fe X-ray emission in Cas A \citep{ghb14}.

\acknowledgments
This research has been supported by the agreement ASI/INAF n.2013-025.R1 and the grant PRIN-2014 (ÒTransient Universe, unveiling new types of stellar explosions with PESSTOÓ).

\begin{deluxetable}{ccccccccccrrrrrr}
\tabletypesize{\scriptsize}
\setlength{\tabcolsep}{0.040in}  
\tablecaption{Abundances of \nuk{Ti}{44} and \nuk{Ni}{56}}
\tablecolumns{16}
\tablenum{1}
\tablewidth{0pt}
\tablehead{
$\rm M_{ini}$ &
Mcut   &
M(Siix)   &
M(Ox)   &
\nuk{Ti}{44}  &
\nuk{Ti}{44}  &
\nuk{Ti}{44}  &
\nuk{Ni}{56}  &
\nuk{Ni}{56}  &
\nuk{Ni}{56}  & 
\nuk{Ti}{44}  &
\nuk{Ti}{44}  &
\nuk{Ti}{44}  &
\nuk{Ni}{56}  &
\nuk{Ni}{56}  &
\nuk{Ni}{56} \\
 &
 &
 &
 &
 Mcut  &
 Siix  &
 Ox    &
 Mcut  &
 Siix  &
 Ox    & 
 Mcut  &
 Siix  &
 Ox    &
 Mcut  &
 Siix  &
 Ox    \\
 $\rm M_\odot$ & $\rm M_\odot$ & $\rm M_\odot$ & $\rm M_\odot$ & $\rm M_\odot$ & $\rm M_\odot$ & $\rm M_\odot$ & $\rm M_\odot$ & $\rm M_\odot$ & $\rm M_\odot$ & $\%$ & $\%$ & $\%$ & $\%$ & $\%$ & $\%$ \\
}
\startdata
\cutinhead{[Fe/H]=0  v=0 km/s}
 13 & 1.52 & 1.59 & 1.65 &   2.6(-05) &  2.2(-06) &  6.6(-07) &  7.0(-02) &  1.2(-02) &  4.4(-05) &   91.7 &   5.7 &   2.6 &  82.6 &  17.3 &  0.1 \\  
 15 & 1.63 & 1.68 & 1.80 &   1.3(-05) &  3.7(-06) &  9.9(-07) &  7.0(-02) &  2.5(-02) &  8.2(-06) &   71.4 &  21.0 &   7.6 &  63.9 &  36.1 &  0.0 \\ 
 20 & 1.62 & 1.67 & 1.79 &   1.5(-05) &  4.0(-06) &  1.5(-06) &  7.0(-02) &  2.9(-02) &  2.7(-04) &   73.1 &  16.5 &  10.4 &  58.0 &  41.6 &  0.4 \\ 
 25 & 1.87 & 1.91 & 2.06 &   8.9(-06) &  5.3(-06) &  1.9(-06) &  7.0(-02) &  4.4(-02) &  6.3(-05) &   39.9 &  38.5 &  21.5 &  37.8 &  62.1 &  0.1 \\
 30 & 2.66 & 2.66 & 3.08 &   1.3(-05) &  1.3(-05) &  3.2(-06) &  7.0(-02) &  7.0(-02) &  1.9(-05) &    0.0 &  75.7 &  24.3 &   0.0 & 100.0 &  0.0 \\
 40 & 2.66 & 2.66 & 2.94 &   1.3(-05) &  1.3(-05) &  6.9(-06) &  7.0(-02) &  7.0(-02) &  1.4(-03) &    0.0 &  48.2 &  51.8 &   0.0 &  98.0 &  2.0 \\
 60 & 3.23 & 3.23 & 3.72 &   1.9(-05) &  1.9(-05) &  8.6(-06) &  7.0(-02) &  7.0(-02) &  1.1(-03) &    0.0 &  54.4 &  45.6 &   0.0 &  98.4 &  1.6 \\
 80 & 4.25 & 4.25 & 4.92 &   3.0(-05) &  3.0(-05) &  1.6(-05) &  7.0(-02) &  7.0(-02) &  1.4(-03) &    0.0 &  47.6 &  52.4 &   0.0 &  98.0 &  2.0 \\
120 & 4.73 & 4.73 & 5.42 &   3.1(-05) &  3.1(-05) &  2.0(-05) &  7.0(-02) &  7.0(-02) &  3.9(-03) &    0.0 &  36.1 &  63.9 &   0.0 &  94.5 &  5.5 \\
\cutinhead{[Fe/H]=0  v=300 km/s}
 13 & 2.02 & 2.05 & 2.27 &   2.6(-05) &  2.4(-05) &  1.8(-05) &  7.0(-02) &  4.2(-02) &  4.8(-04) &    8.3 &  22.0 &  69.6 &  39.4 &  59.9 &  0.7 \\
 15 & 2.30 & 2.35 & 2.77 &   6.0(-06) &  5.9(-06) &  2.1(-06) &  7.0(-02) &  4.4(-02) &  1.6(-06) &    1.7 &  63.6 &  34.7 &  37.3 &  62.7 &  0.0 \\
 20 & 2.72 & 2.72 & 3.17 &   1.3(-05) &  1.3(-05) &  3.9(-06) &  7.0(-02) &  7.0(-02) &  1.3(-05) &    0.0 &  70.5 &  29.5 &   0.0 & 100.0 &  0.0 \\
 25 & 2.49 & 2.49 & 2.86 &   1.4(-05) &  1.4(-05) &  4.9(-06) &  7.0(-02) &  7.0(-02) &  4.1(-04) &    0.0 &  63.8 &  36.2 &   0.0 &  99.4 &  0.6 \\
 30 & 2.36 & 2.36 & 2.63 &   1.3(-05) &  1.3(-05) &  6.3(-06) &  7.0(-02) &  7.0(-02) &  1.0(-03) &    0.0 &  51.3 &  48.7 &   0.0 &  98.5 &  1.5 \\
 40 & 2.95 & 2.95 & 3.27 &   1.7(-05) &  1.7(-05) &  1.1(-05) &  7.0(-02) &  7.0(-02) &  2.7(-03) &    0.0 &  38.9 &  61.1 &   0.0 &  96.1 &  3.9 \\
 60 & 3.30 & 3.30 & 3.74 &   1.9(-05) &  1.9(-05) &  1.0(-05) &  7.0(-02) &  7.0(-02) &  3.0(-03) &    0.0 &  44.6 &  55.4 &   0.0 &  95.7 &  4.3 \\
 80 & 3.47 & 3.47 & 3.95 &   2.1(-05) &  2.1(-05) &  1.2(-05) &  7.0(-02) &  7.0(-02) &  3.2(-03) &    0.0 &  41.6 &  58.4 &   0.0 &  95.5 &  4.5 \\
120 & 3.49 & 3.49 & 4.11 &   2.1(-05) &  2.1(-05) &  6.4(-06) &  7.0(-02) &  7.0(-02) &  1.1(-05) &    0.0 &  68.8 &  31.2 &   0.0 & 100.0 &  0.0 \\
\cutinhead{[Fe/H]=-1  v=0 km/s}
 13 & 1.53 & 1.60 & 1.66 &   2.3(-05) &  2.6(-06) &  8.1(-07) &  7.0(-02) &  1.5(-02) &  5.1(-05) &   88.7 &   7.8 &   3.5 &  78.2 &  21.7 &  0.1 \\
 15 & 1.69 & 1.72 & 1.91 &   9.1(-06) &  6.3(-06) &  1.8(-06) &  7.0(-02) &  4.4(-02) &  1.3(-05) &   31.3 &  49.5 &  19.2 &  36.8 &  63.2 &  0.0 \\
 20 & 1.74 & 1.78 & 1.92 &   9.9(-06) &  4.3(-06) &  1.1(-06) &  7.0(-02) &  3.5(-02) &  4.4(-06) &   56.9 &  31.9 &  11.2 &  50.3 &  49.7 &  0.0 \\
 25 & 2.48 & 2.48 & 2.84 &   1.3(-05) &  1.3(-05) &  3.4(-06) &  7.0(-02) &  7.0(-02) &  8.0(-05) &    0.0 &  74.2 &  25.8 &   0.0 &  99.9 &  0.1 \\
 30 & 2.52 & 2.52 & 2.91 &   1.4(-05) &  1.4(-05) &  3.2(-06) &  7.0(-02) &  7.0(-02) &  3.1(-05) &    0.0 &  77.4 &  22.6 &   0.0 & 100.0 &  0.0 \\
 40 & 2.65 & 2.65 & 2.97 &   1.6(-05) &  1.6(-05) &  6.0(-06) &  7.0(-02) &  7.0(-02) &  4.5(-04) &    0.0 &  63.7 &  36.3 &   0.0 &  99.4 &  0.6 \\
 60 & 4.48 & 4.48 & 5.10 &   3.9(-05) &  3.9(-05) &  1.8(-05) &  7.0(-02) &  7.0(-02) &  9.0(-04) &    0.0 &  53.6 &  46.4 &   0.0 &  98.7 &  1.3 \\
 80 & 6.96 & 6.96 & 7.70 &   5.3(-05) &  5.3(-05) &  2.5(-05) &  7.0(-02) &  7.0(-02) &  7.1(-04) &    0.0 &  52.5 &  47.5 &   0.0 &  99.0 &  1.0 \\
120 & 5.73 & 5.73 & 6.37 &   6.0(-05) &  6.0(-05) &  4.6(-05) &  7.0(-02) &  7.0(-02) &  9.1(-03) &    0.0 &  23.5 &  76.5 &   0.0 &  87.0 & 13.0 \\
\cutinhead{[Fe/H]=-1  v=300 km/s}
 13 & 2.10 & 2.13 & 2.36 &   3.9(-05) &  3.4(-05) &  2.9(-05) &  7.0(-02) &  5.0(-02) &  6.2(-05) &   12.5 &  12.8 &  74.7 &  28.3 &  71.6 &  0.1 \\
 15 & 2.23 & 2.27 & 2.47 &   4.6(-05) &  4.5(-05) &  4.1(-05) &  7.0(-02) &  3.8(-02) &  1.1(-03) &    2.1 &   8.3 &  89.6 &  45.3 &  53.1 &  1.5 \\
 20 & 2.62 & 2.62 & 2.98 &   1.3(-05) &  1.3(-05) &  5.7(-06) &  7.0(-02) &  7.0(-02) &  1.5(-03) &    0.0 &  55.0 &  45.0 &   0.0 &  97.9 &  2.1 \\
 25 & 2.89 & 2.89 & 3.26 &   2.0(-05) &  2.0(-05) &  8.8(-06) &  7.0(-02) &  7.0(-02) &  8.6(-04) &    0.0 &  54.9 &  45.1 &   0.0 &  98.8 &  1.2 \\
 30 & 3.70 & 3.70 & 4.19 &   2.6(-05) &  2.6(-05) &  1.2(-05) &  7.0(-02) &  7.0(-02) &  8.6(-04) &    0.0 &  55.3 &  44.7 &   0.0 &  98.8 &  1.2 \\
 40 & 4.28 & 4.28 & 4.84 &   3.1(-05) &  3.1(-05) &  1.6(-05) &  7.0(-02) &  7.0(-02) &  1.2(-03) &    0.0 &  49.3 &  50.7 &   0.0 &  98.3 &  1.7 \\
 60 & 5.05 & 5.05 & 5.63 &   3.8(-05) &  3.8(-05) &  2.4(-05) &  7.0(-02) &  7.0(-02) &  4.3(-03) &    0.0 &  36.5 &  63.5 &   0.0 &  93.9 &  6.1 \\
 80 & 5.76 & 5.76 & 6.28 &   4.4(-05) &  4.4(-05) &  3.3(-05) &  7.0(-02) &  7.0(-02) &  8.4(-03) &    0.0 &  25.4 &  74.6 &   0.0 &  88.0 & 12.0 \\
120 & 8.01 & 8.01 & 8.61 &   5.8(-05) &  5.8(-05) &  4.5(-05) &  7.0(-02) &  7.0(-02) &  8.6(-03) &    0.0 &  23.6 &  76.4 &   0.0 &  87.7 & 12.3 \\
\enddata
\end{deluxetable}

\end{document}